\documentclass[12pt]{article}
\usepackage{latexsym}
\usepackage{amssymb}
\usepackage{graphicx}
\usepackage{multibox}
\usepackage{amsmath,amsfonts}

\topskip \textwidth 480pt

\if{0}
\documentclass[fdp,a4paper,fleqn%
]{w-art}
\usepackage{times,cite,w-thm}
\theoremstyle{plain}

\theoremstyle{definition}

\usepackage[]{graphicx}
\fi
\begin{document}

\if{0}
\DOIsuffix{theDOIsuffix}
\Volume{55}
\Month{01}
\Year{2007}
\pagespan{1}{}

\Receiveddate{XXXX}
\Reviseddate{XXXX}
\Accepteddate{XXXX}
\Dateposted{XXXX}
\fi
\begin{titlepage}
\title{
\begin{flushright}
{\small MIFPA-11-01}\\
~\\
~\\
\end{flushright}
{\bf Peculiarities of String Theory on $AdS_4 \times CP^3$}
~\\
\medskip
\medskip
\medskip
\medskip
\author{Dmitri~Sorokin\footnote{\tt dmitri.sorokin@pd.infn.it}$\,$ and Linus Wulff${~}^*$\footnote{\tt linus@physics.tamu.edu}
~\\
~\\
~\\
{$^*$\it Istituto Nazionale di Fisica Nucleare, Sezione di Padova,}
~\\
{\it via F. Marzolo 8, 35131 Padova, Italia}
~\\
~\\
{$^\dagger$\it George and Cynthia Woods Mitchell Institute}
~\\
{\it for Fundamental Physics and Astronomy,}
~\\
{\it Texas A\&M University, College Station, TX 77843, USA} 
}}

\date{}
\maketitle

\begin{abstract}
We review peculiar features of type IIA string theory compactified on $AdS_4\times
CP^3$ superspace, in particular, the structure of the Green--Schwarz action, issues with fixing its kappa--symmetry, classical integrability and the string instanton on $CP^3$.

\end{abstract}

\thispagestyle{empty}
\end{titlepage}

\section{Introduction}
The present interest in type IIA string theory on the $AdS_4\times CP^3$ background is mainly motivated by the $AdS_4/CFT_3$ correspondence. The knowledge of the structure of the worldsheet theory is necessary for working on various problems of this correspondence from the string theory side. In this contribution we discuss some peculiar features of the string theory on $AdS_4\times CP^3$ which make it different from the prominent $AdS_5 \times S^5$ superstring. First of all, obviously, the target spaces of the two theories are different, but, probably, the most crucial difference is that $AdS_4\times CP^3$ theory is less supersymmetric. It possesses 24 of 32 type IIA D=10 supersymmetries which, together with the $AdS_4\times CP^3$ isometries, form the supergroup $OSp(6|4)$.

The lack of supersymmetries causes a more complicated structure of the $AdS_4\times CP^3$ superspace in which the string propagates. Though it has the $OSp(6|4)$ isometry and its bosonic subspace is the direct product of symmetric spaces ($AdS_4=SO(2,3)/SO(1,3)$ and $CP^3=SU(4)/U(3)$), this superspace (with ten bosonic and thirty two fermionic directions) is not a supercoset manifold. This is in contrast to the $AdS_5 \times S^5$ case in which the background superspace is the supercoset $PSU(2,2|4)/SO(1,4)\times SO(5)$.

The more complicated, non--supercoset, structure of the $AdS_4\times CP^3$ superspace makes the proof of the classical integrability of the worldsheet theory in this target superspace much more tricky than in the $AdS_5 \times S^5$ case.

Lack of some supersymmetry can, perhaps, be a reason that some basic issues have still not been completely settled on both sides of the $AdS_4/CFT_3$ correspondence (see \emph{e.g.} \cite{Klose:2010ki} for a recent review). For instance, there are still some subtleties and discrepancies in matching  $CFT_3$ anomalous dimensions with string energies (see \emph{ e.g.} \cite{Gaiotto:2008cg,Nishioka:2008gz,Grignani:2008is,Grignani:2008te,Astolfi:2008ji,McLoughlin:2008ms,Alday:2008ut,Krishnan:2008zs,McLoughlin:2008he,Chen:2008qq,Gromov:2008fy,Bandres:2009kw,Abbott:2010yb,Astolfi:2011ju}). There are subtleties in matching worldsheet degrees of freedom with those of S-matrix scattering theory \cite{Zarembo:2009au} (there are light and heavy worldsheet modes, while the scattering theory describes only light ones \cite{Ahn:2008aa}). There is an issue regarding the existence of dual superconformal symmetry in the ABJM model \cite{Bargheer:2010hn,Henn:2010ps,Lee:2010du,Huang:2010qy,Gang:2010gy} and the fermionic T--duality of the $AdS_4\times CP^3$ superstring \cite{Adam:2009kt,Grassi:2009yj,Hao:2009hw,Adam:2010hh,Bakhmatov:2010fp,Dekel:2011qw}. The $AdS_5/CFT_4$ counterparts of these dual symmetries have been known to play an important role in revealing and understanding various properties of the  correspondence (see \emph{e.g.} \cite{Alday:2007hr,Drummond:2007aua,Drummond:2007cf,Brandhuber:2007yx,Drummond:2008vq,Komargodski:2008wa,Ricci:2007eq,Berkovits:2008ic,Beisert:2008iq}).

Another peculiarity of the $AdS_4\times CP^3$ string theory is that it has worldsheet instantons which wrap a non--trivial 2--cycle in $CP^3$ \cite{Cagnazzo:2009zh}. In contrast, there are no worldsheet instantons in string theory on $AdS_5\times S^5$ but there are $D$--instantons. The role of the latter in the $AdS_5/CFT_4$ correspondence is understood pretty well, their CFT duals being the conventional gauge--field instantons (see \emph{ e.g.} \cite{Bianchi:2007ft} for a review and references). The role of the worldsheet instantons in the $AdS_4/CFT_3$ holography should still be unveiled (for first results in this direction see \cite{Bianchi:2010mg,Drukker:2010nc,Bianchi:2010aw}). Note also that on the CFT side of the correspondence there exist monopole instantons in the ABJM model \cite{Hosomichi:2008ip} which, however, do not have (at least a direct) relation to the $AdS_4\times CP^3$ string instantons.

In what follows we shall discuss the structure of the Green--Schwarz action for the superstring on $AdS_4\times CP^3$ \cite{Gomis:2008jt}, subtleties of its kappa--symmetry gauge fixing and reduction to the $OSp(6|4)/SO(1,3)\times U(3)$ supercoset sigma--model of \cite{Arutyunov:2008if,Stefanski:2008ik,Fre:2008qc}, the evidence for the classical integrability of this theory \cite{Arutyunov:2008if,Stefanski:2008ik,Sorokin:2010wn} and properties of the worldsheet instanton on $CP^3$ \cite{Cagnazzo:2009zh}.

\section{Green--Schwarz superstrings in a generic supergravity background}
Let us start by recalling the basics of the Green--Schwarz formulation of the superstring. In this formulation the string propagates in a (generally curved) superspace parametrized by ten bosonic coordinates $X^M$ ($M=0,1,\cdots,9)$ and thirty two fermionic coordinates $\Theta^{\underline\mu}$ ($\underline\mu=1,\cdots,32$), which we shall collectively call $Z^{\mathcal M}=(X^M,\Theta^{\underline\mu})$. We are interested in type IIA theory in which $\Theta$ are associated with a $D=10$ Majorana spinor. As usual, the geometry of the superspace is described by frame one--forms, or supervielbeins,
\begin{equation}\label{E}
E^A(X,\Theta)=dZ^{\mathcal M}E_{\mathcal M}{}^A\,, \qquad E^{\underline\alpha}(X,\Theta)=dZ^{\mathcal M}E_{\mathcal M}{}^{\underline\alpha}
\end{equation}
and by a spin connection one--form
\begin{equation}\label{Omega}
\Omega_{A}{}^{B}(X,\Theta)=dZ^{\mathcal M}\Omega_{\mathcal M\,A}{}^B\,.
\end{equation}
The vielbein components $E^A$ point along the bosonic tangent--space directions and $E^{\underline\alpha}$ point along the fermionic tangent--space directions. To describe supergravity the supervielbeins should be constrained. The basic (and the most essential) constraint (which is common to all supergravity theories) is the torsion constraint
\begin{equation}\label{TA}
T^A:=dE^A+E^B\Omega_B{}^A=-iE^{\underline\alpha}\Gamma^A_{\underline{\alpha\beta}}E^{\underline\beta}\,,
\end{equation}
where $\Gamma^A_{\underline{\alpha\beta}}$ are $D=10$ gamma--matrices. The geometrical meaning of this constraint is that in the flat--space limit it ensures the superspace to have the geometry of conventional Minkowski superspace with a constant torsion.

An important physical meaning of this constraint is that it reduces the number of independent component fields contained in the superfields $E^{\mathcal A}$ and $\Omega_B{}^A$, and makes their contents to be that of the supergravity multiplet. In addition, in the case of \emph{e.g.} type II $D=10$ supergravities this constraint implies the equations of motion for the supergravity component fields, which one recovers by studying the consistency of this constraint with the Bianchi identities on torsion and curvature. In other words the constraint \eqref{TA} puts type IIA supergravity on the mass shell and allows one to derive the dependence of the superfields $E^{\mathcal A}$ and $\Omega_B{}^A$ on the physical component fields of the theory. For instance, expanding the vector supervielbein $E^A$ in powers of $\Theta$ we get (schematically)
\begin{eqnarray}\label{eE}
E_M{}^A(X,\Theta)&=&e_M{}^A(X)+(\psi_M+\lambda\Gamma_M)\Gamma^A\Theta+\omega_M{}^{BC}\Theta\Gamma^A{}_{BC}\Theta\\
&&\hspace{-100pt}+ H_{MBC}\Theta\Gamma^{ABC}\Gamma_{11}\Theta+ e^{\phi}\,F_{BC}\Theta\Gamma^A\Gamma^{BC}\Gamma_M\Gamma_{11}\Theta+e^{\phi}F_{BCDK}\Theta\Gamma^A\Gamma^{BCDK}\Gamma_M\Theta+\cdots\,,
\nonumber\end{eqnarray}
where we recognize members of the type IIA supergravity multiplet: the graviton $e_M{}^A(X)$, gravitino $\psi_M(X)$, dilatino $\lambda(X)$, dilaton $\phi(X)$, the NS--NS field strength $H_3=dB_2(X)$ and the RR field strengths $F_2=dA_1(X)$ and $F_4=dA_3(X)+A_1 H_3(X)$.

We are now ready to present the Green--Schwarz superstring action. It consists of two terms, the Nambu--Goto term and the Wess--Zumino term,
\begin{equation}\label{cordaA}
S = -T\,\int d^2\xi\, \sqrt {-\det{{E}_i{}^A {E}_j{}^B \eta_{AB}}}\,
-T\,\int  B_2(X,\Theta)\,,
\end{equation}
where $T=\frac{1}{2\pi\alpha'}$ is the string tension, $\xi^i$  $(i,j=0,1)$ are the worldsheet coordinates, ${E_i{}^A}=\partial_iZ^{\mathcal M}E_{\mathcal M}{}^A(Z)$ are worldsheet pullbacks of the target superspace vector supervielbeins and $B_2(X,\Theta)$ is the pull--back of the NS--NS 2--form superfield.

Via the superfields $E^A(X,\Theta)$ and $B_2(X,\Theta)$ the string couples to all component fields of the supergravity multiplet including the RR fields, which are encoded in the form of the superfields like eq. \eqref{eE}. This makes the Green--Schwarz formulation indispensable to the study of the string in Ramond--Ramond backgrounds, since the classical Ramond--Neveu--Schwarz formulation does not include the RR fields. So if one knows the explicit form of $E^A(X,\Theta)$ and $B_2(X,\Theta)$ in terms of the component fields to all (thirty two) orders in $\Theta$ then one knows the explicit form of string couplings to supergravity.

The Green--Schwarz formulation has an important local fermionic symmetry, the kappa--symmetry, which reflects the existence of supersymmetric BPS configurations of strings (and branes) and which in some cases allows one to considerably simplify the structure of the action by removing pure--gauge fermionic degrees of freedom.
The kappa--symmetry transformations  of the worldsheet fields
$Z^{\mathcal M}(\xi)=(X^M(\xi),\Theta^{\underline \mu}(\xi))$
leave the superstring action (\ref{cordaA}) invariant provided the superbackground obeys the superspace supergravity constraints like \eqref{TA}. The kappa--variations of $Z^{\mathcal M}$ are such that they are zero along the tangent--space vector directions
\begin{equation}\label{kA}
\delta_\kappa Z^{\mathcal M}\,{E}_{\mathcal M}{}^A=0,
\qquad
\end{equation}
while along the spinorial directions they have the following form
\begin{equation}\label{kappastring}
\delta_\kappa Z^{\mathcal M}\,{E}_{\mathcal M}{}^{\underline \alpha}=
\frac{1}{2}(1+\Gamma)^{\underline \alpha}_{~\underline\beta}\,
\kappa^{\underline\beta}(\xi)\,,
\end{equation}
where $\kappa^{\underline\alpha}(\xi)$ is a 32--component Grassmann--odd spinor
parameter, $\frac{1}{2}(1+\Gamma)^{\underline
\alpha}_{~\underline\beta}$ is a spinor projection matrix with
\begin{equation}\label{gbs}
\Gamma=\frac{1}{2\,\sqrt{-\det{G}}}\,\varepsilon^{ij}\,E_i{}^A\,E_j{}^B\,\Gamma_{AB}\,\Gamma_{11},\qquad (\Gamma)^2=1\,
\end{equation}
where $\varepsilon^{01}=-\varepsilon^{10}=1$ and $G_{ij}={E}_i{}^A\,{E}_j{}^B\,\eta_{AB}$ is
the induced metric on the worldsheet.

Due to the presence of the projector, only half of the components of $\kappa^{\underline\alpha}(\xi)$ are involved in the kappa--transformations. Thus, kappa--symmetry can be used to gauge away 16 of the 32 worldsheet fermionic modes $\Theta^{\underline\mu}(\xi)$. Upon having reviewed generic properties of the Green--Schwarz formulation, we are now in a position to discuss the main features of the $AdS_4\times CP^3$ superstring.

\section{String in the $AdS_4\times CP^3$ superbackground}
The $AdS_4\times CP^3$ solution of type IIA supergravity has been known since the early 80s \cite{Watamura:1983hj} and was studied in detail by Nilsson and Pope \cite{Nilsson:1984bj}. The compactification of the $D=10$ space--time into $AdS_4\times CP^3$ is carried out by non--zero expectation values of the $F_2$ and $F_4$ RR fluxes. The $F_4$ flux is proportional to the volume four--form on $AdS_4$
\begin{equation}\label{f4}
F_4=\frac{3e^{-\phi}}{R_{AdS}}\,\sqrt{-g_{AdS}}\,dx^0dx^1dx^2dx^3 \quad \Rightarrow\quad F_{abcd}=-\frac{3e^{-\phi}}{R_{AdS}}\,\varepsilon_{abcd}\,, \qquad \varepsilon_{0123}=-1
\end{equation}
and the $F_2$ flux is proportional to the K\"ahler form on $CP^3$
\begin{equation}\label{f2}
F_2=-\frac{e^{-\phi}}{R_{AdS}}\,J_2 \qquad \Rightarrow\qquad F_{a'b'}=-\frac{e^{-\phi}}{R_{AdS}}\,J_{a'b'}\,
\end{equation}
where $a,b,\ldots=0,\ldots,3$ and $a',b',\ldots=4,\ldots,9$ are $AdS_4$ and $CP^3$ tangent space indices respectively, $R_{AdS}$ is the $AdS_4$ radius which is half of the $CP^3$ one and $\phi$ is a constant dilaton which defines the string coupling constant $g_{str}$ and is related to the rank $N$ of the gauge group, the Chern--Simons level $k$ and the 't Hooft coupling constant $\lambda=N/k$ of the dual $CFT_3$ theory as follows \cite{Aharony:2008ug}
$$
e^{2\phi}=g^2_{str}=\frac{\lambda^{5/2}}{N^2}, \qquad \lambda=\frac{N}{k}.
$$
Note also that in string units $R^2_{AdS}=\sqrt{2\pi^2\,\lambda}\,\alpha'$ \footnote{From the above relations between the parameters one deduces that
the Type IIA supergravity regime is valid when $\lambda \gg 1$, since in string units  $R^2_{AdS} \sim \sqrt{\lambda}\cdot\alpha'$ and is large when $\lambda \gg 1$. The string theory regime is valid when $\lambda^{5/2}\gg{N^2}$, while the $D=11$ supergravity approximation is valid when $N\gg k^5$, since in Plank units  the AdS radius is $R^6_{AdS} \sim Nk$, but the $S^1$ radius is $R/k$ and is large when $N>>k^5$. The relation between $l_p$ and $\alpha'$ is $l_p = (g_{str})^{1/3}(\alpha')^{1/2}$.}.

Our goal is to understand how the string interacts with the $AdS_4\times CP^3$ background and, in particular, with the $F_2$ and $F_4$ fluxes. As we have seen, the supergravity fields enter the Green--Schwarz action as the component fields in the $\Theta$--expansion of the supervielbeins $E^A(X,\Theta)$ and the NS--NS two--form $B_2(X,\Theta)$. Therefore, to know the explicit form of the Green--Schwarz action we should derive an explicit form of the type IIA superfields to all 32 orders in $\Theta$. In principle, this can be done by solving the superfield supergravity constraints, like \eqref{TA}, and corresponding Bianchi identities, taking the bosonic $AdS_4\times CP^3$ background as the initial condition. However, such a computation is very tedious and, in general, can hardly lead to a reasonably treatable result due to the complexity. For instance, so far D. Tsimpis \cite{Tsimpis:2004gq} has managed to arrive at the 5-th order in $\Theta$ in the generic case of $D=11$ supergravity. On the other hand, when the background possesses a huge symmetry and is maximally supersymmetric, as in the case of $AdS_4\times S^7$ and $AdS_7\times S^4$ in $D=11$, or $AdS_5\times S^5$ in type IIB string theory, the corresponding superspace geometry turns out to be that of supercoset spaces whose bosonic subspaces are the backgrounds of interest. For example, the $AdS_5\times S^5$ superspace is the supercoset $PSU(2,2|4)/SO(1,4)\times SO(5)$ and the structure of its supervielbeins and gauge--field superforms can be derived using the Maurer--Cartan equations in an explicit compact form (see e.g. \cite{Metsaev:1998it,Kallosh:1998zx}).

So the natural thing to do also in the $AdS_4\times CP^3$ case is to try to reduce the string action to a sigma model on the supercoset space $OSp(6|4)/SO(1,3)\times U(3)$ which contains $AdS_4\times CP^3$ as the bosonic subspace but has only 24 fermionic directions. Such a model was constructed by several groups of authors \cite{Arutyunov:2008if,Stefanski:2008ik,Fre:2008qc,Bonelli:2008us,D'Auria:2008cw}. It was assumed that this supercoset model is a partially gauge--fixed version of the complete Green--Schwarz superstring in which the eight fermions corresponding to broken supersymmetries were gauged away using kappa symmetry. To identify these fermions we should understand which 24 supersymmetries of the 32 ones get preserved by the $AdS_4\times CP^3$ background. To this end let us look at the supersymmetry transformations of the fermionic fields, i.e. gravitino and dilatino.  Since in the background under consideration the fermionic fields are zero, the background is invariant under those supersymmetries under which the fermionic fields remain zero. In particular, consider the variation of the type IIA dilatino that has the following form (when the dilaton is constant and the NS--NS flux $H_3$ is zero)
\begin{equation}\label{susyvar}
\delta\lambda=\frac{e^\phi}{8}\Big(\frac{1}{4!}\,F_{MNPQ}\Gamma^{MNPQ}+\frac{3}{2}\,F_{MN}\Gamma^{MN}\Gamma_{11}\Big)\epsilon\,.
\end{equation}
In our case the RR fluxes are those in eqs. \eqref{f4} and \eqref{f2}, so the variation of the dilatino reduces to
\begin{equation}\label{susyvar1}
\delta\lambda=-\frac{3i}{16R_{AdS}}\,\gamma_5\Big(2-iJ_{a'b'}\Gamma^{a'b'}\gamma_{7}\Big)\epsilon\,,
\end{equation}
where $\gamma_5=i\Gamma^{0123}$ and $\gamma_7=i\Gamma^{456789}$. We see that the variation is zero if $\epsilon$ satisfy the condition
\begin{equation}\label{susycon}
\frac{1}{8}\Big(2-iJ_{a'b'}\Gamma^{a'b'}\gamma_{7}\Big)\epsilon\equiv \mathcal P_8\,\epsilon=0.
\end{equation}
It turns out that $\mathcal P_8$ is a $32\times 32$ projection matrix which has eight non--zero eigenvalues \cite{Nilsson:1984bj}. Thus it annihilates 24 of the 32 components of $\epsilon$, and these 24 components correspond to the unbroken supersymmetries.

In the superspace formulation,  supersymmetry transformations act (at the linearized level) as translations on the fermionic coordinates of the superspace
$$
\Theta'=\Theta+\epsilon.
$$
Therefore, the 24 fermionic coordinates satisfying the projection condition
\begin{equation}\label{vartheta}
\mathcal P_8\Theta=0 \qquad\Rightarrow\qquad \vartheta\equiv (1-\mathcal P_8)\Theta =\mathcal P_{24}\Theta
\end{equation}
are associated with the unbroken supersymmetries, while the remaining eight coordinates
\begin{equation}\label{upsilon}
\upsilon=\mathcal P_8\Theta
\end{equation}
are associated with the broken supersymmetries. It is natural to try to get rid of these coordinates in the superstring action by using kappa--symmetry \eqref{kappastring}. If it is possible, we are left with ten bosonic coordinates, $X^M=(x^m,y^{m'})$, of $AdS_4\times CP^3$ and 24 fermionic coordinates $\vartheta$ \eqref{vartheta}. These are what we need for parametrizing the supercoset space $OSp(6|4)/U(3)\times SO(1,3)$. Then one can construct a worldsheet action on this supercoset, similar to the supercoset sigma--model for the $AdS_5\times S^5$ superstring. The building blocks for this action are the components of the Cartan form
\begin{equation}\label{cartan}
K^{-1}\,dK(x,y,\vartheta)=E^a(x,y,\vartheta)\,P_a+E^{a'}(x,y,\vartheta)\,P_{a'}+E^{\alpha a'}(x,y,\vartheta)\,Q_{\alpha a'}
+\Omega(x,y,\vartheta)\,M\,
\end{equation}
where $K(x,y,\vartheta)$ is a coset element of the supergroup $OSp(6|4)$, and $P_a$, $P_{a'}$, $Q_{\alpha\alpha'}$ and $M$ are the generators of the $OSp(6|4)$ algebra. The indices $\alpha$ label the $AdS_4$ Majorana spinors and the indices $a'=1,\cdots 6$ label the six--dimensional representation of $SU(3)$. The Cartan form component $\Omega$ associated with the generators $M$ of the stability subgroup $U(3)\times SO(1,3)$ plays the role of the spin connection on this superspace. The Cartan form components associated with the generators $P_a$ and $P_{a'}$ of translations along $AdS_4\times CP^3$ are vector supervielbeins of this superspace whose pullbacks are used to construct the Nambu--Goto part of the Green--Schwarz action \eqref{cordaA}, while the fermionic supervielbeins $E^{\alpha a'}(x,y,\vartheta)$ can be used to construct its Wess--Zumino term in a simple explicit form \cite{Berkovits:1999zq}. The $OSp(6|4)/U(3)\times SO(1,3)$ sigma--model action has the following form \cite{Arutyunov:2008if,Stefanski:2008ik}
\begin{equation}\label{sigmaction}
S=-T\int d^2\xi\sqrt{-\det {E_i{}^AE_j{}^B}\,\eta_{AB}}-T\int E^{\alpha a'}\wedge E^{\beta b'}\,J_{a'b'}\,C_{\alpha\beta}\,,
\end{equation}
where we have given the first term in the Nambu--Goto form instead of making use of an auxiliary worldsheet metric and $C_{\alpha\beta}=-C_{\beta\alpha}$ is the $D=4$ charge conjugation matrix.
Let us note that the functional dependence of $E^A(X,\vartheta)$ and $E^{\alpha a'}(X,\vartheta)$ can be derived in an explicit form using a chosen realization of the coset element $K(X,\vartheta)$, e.g. an exponential one $K(X,\vartheta)=e^{XP}e^{\vartheta Q}$.

By analogy, with the $AdS_5\times S^5$ supercoset sigma--model \cite{Bena:2003wd}, the sigma--model \eqref{sigmaction} was shown to be classically integrable \cite{Arutyunov:2008if,Stefanski:2008ik}. However, it does not describe all possible sectors of string theory in $AdS_4\times CP^3$. Only when the superstring
is extended in $CP^3$ can its dynamics be described by this $OSp(6|4)/U(3)\times SO(1,3)$ supercoset sigma--model
\cite{Arutyunov:2008if,Gomis:2008jt}. The reason is that the kappa--symmetry gauge fixing which we used to eliminate the eight broken supersymmetry fermions $\upsilon$ is not always admissible. For instance, it is  not admissible when the string moves entirely in the $AdS_4$ part of the superbackground or forms a worldsheet instanton in $CP^3$. To verify that kappa--symmetry can eliminate all the eight $\upsilon$, one should check that for a given classical solution the kappa--symmetry projector \eqref{gbs} does not commute with the projector $\mathcal P_8$ \eqref{susycon}. It is not hard to see that, e.g. when the string moves in $AdS_4$ only (i.e. the $CP^3$ coordinates $y^{m'}$ are worldsheet constants) the two projectors commute and this means that kappa--symmetry can eliminate only half of $\upsilon$.  In the supercoset sigma--model this problem manifests itself in the fact that in the $AdS_4$ subsector the number of independent kappa--symmetries gets increased from eight to twelve which results in the loss of four physical fermionic degrees of freedom.  Therefore, to describe the string theory in such singular sectors one needs to know the form of the Green--Schwarz action in the $AdS_4\times CP^3$ superspace with all 32 fermionic directions.

The simplest way to get the geometrical structure of this superspace turned out to be the Kaluza--Klein dimensional reduction of the maximally supersymmetric $AdS_4\times S^7$ solution of $D=11$ supergravity whose superspace is the supercoset manifold
$OSp(8|4)/SO(7)\times SO(1,3)$. It has been known since the early 80s that the type IIA $AdS_4\times CP^3$ solution is related to the $D=11$ $AdS_4\times S^7$ solution by dimensional reduction \cite{Nilsson:1984bj,Sorokin:1984ca,Sorokin:1985ap}. The geometrical ground for this relation is the Hopf fibration structure of $S^7$ with $CP^3$ being the base and an $S^1\sim U(1)$ circle being the fiber. So, to derive the geometry of the $AdS_4\times CP^3$ superspace from the structure of $OSp(8|4)/SO(7)\times SO(1,3)$ one should generalize the Hopf fibration of $S^7$ to the whole superspace $OSp(8|4)/SO(7)\times SO(1,3)$ and then to perform the dimensional reduction of its $U(1)$ fiber. In other words, a key point is to find a parametrization of the $OSp(8|4)/SO(7)\times SO(1,3)$ supercoset geometry which would manifest its structure as a $U(1)$--fiber bundle over a base superspace $\mathcal M_{10,32}$ having $AdS_4\times CP^3$ as the $D=10$ bosonic subspace and 32 fermionic directions. This $\mathcal M_{10,32}$ is the superspace we are looking for. It was constructed, using the above reasoning, in \cite{Gomis:2008jt}. Having at hand the supervielbeins of $\mathcal M_{10,32}$ one plugs them into to the Green--Schwarz action \eqref{cordaA} which can now be used for studying the $AdS_4\times CP^3$ superstring in those sectors of the theory which are not reachable by the supercoset model. For instance, one can try to extend the proof of the classical integrability of the $OSp(6|4)/U(3)\times SO(1,3)$ sigma--model to the complete superstring theory in $AdS_4\times CP^3$.

\section{Integrability of the $AdS_4\times CP^3$ superstring}

The classical integrability of a two--dimensional dynamical system manifests itself in the presence in the theory of an infinite number of conserved currents and charges. These conserved quantities can be derived using a so--called Lax pair or connection. This is a one--form $\mathcal L=d\xi^i\mathcal L_i(\xi,z)$ on the $2d$ worldsheet which takes values in a symmetry algebra, depends on a numerical (spectral) parameter $z$ and has zero curvature
\begin{equation}\label{flatL}
d\mathcal L-\mathcal L\wedge \mathcal L=0\,,
\end{equation}
provided that the equations of motion of the dynamical system are satisfied. And vice versa, the zero--curvature condition should imply the equations of motion. The classical integrability is proven if one manages to construct $\mathcal L(\xi,z)$. However, no generic prescription exists for how to do this. Different systems may require different techniques. In the case of supercoset sigma--models whose isometry supergroup possesses $\mathbf Z_4$--grading, as in the cases of $PSU(2,2|4)/SO(1,4)\times SO(5)$ and $OSp(6|4)/U(3)\times SO(1,3)$, an elegant construction was proposed in \cite{Bena:2003wd}. Without going into details which the reader may find in \cite{Bena:2003wd,Arutyunov:2008if,Stefanski:2008ik}, we present the result of the construction of a Lax connection for the $OSp(6|4)/U(3)\times SO(1,3)$ sigma--model \eqref{cartan} and \eqref{sigmaction}
\begin{equation}\label{L}
\mathcal L(X,\vartheta,z)=\Omega\,M+(l_1E^A +l_2*E^A)P_A +l_3E^{\alpha a'}\,Q_{\alpha a'}
+l_4E^{\alpha a'}\gamma^5_{\alpha\beta}J_{a'b'}\,Q^{\beta b'},
\end{equation}
where the numerical parameters $l_1(z),\cdots, \,l_4(z)$ are restricted by the zero--curvature condition \eqref{flatL} to depend on the single spectral parameter. Note that $\mathcal L$ takes values in the $OSp(6|4)$ algebra.

In the case of the complete Green--Schwarz string (\emph{i.e.} when the kappa--symmetry is not
fixed at all) the superstring moves in $AdS_4\times CP^3$ superspace with thirty two Grassmann--odd
directions and the eight worldsheet fermionic fields associated to the broken supersymmetry
contribute to the structure of the supervielbeins $E^A$, $E^{\alpha a'}$  and to the connection
$\Omega$ thus spoiling their nature as components of the $G/H$ Cartan form. As a result, as one can check by
direct calculations, the $OSp(6|4)$ Lax connection of the form
\eqref{L} constructed from $\Omega$, $E^A$ and $E^{\underline\alpha}$ which include the dependence on these eight
fermions will not have zero curvature for any non--trivial choice of the coefficients. Therefore, a
modification of  \eqref{L} by additional terms depending on the fermions $\upsilon$ is required for
restoring the zero curvature condition \eqref{flatL}.

To reveal the structure of the additional terms, an alternative way of constructing the Lax connection proved to be successful \cite{Sorokin:2010wn}. It uses the structure of the conserved Noether current of the $OSp(6|4)$ symmetry of the complete $AdS_4\times CP^3$ superstring action. The Noether current (written as a $2d$ one--form) is the sum of two currents
\begin{equation}\label{NC}
J=d\xi^iJ_i(X,\vartheta,\upsilon)=J_{\mathcal B}(X,\vartheta,\upsilon)+J_{susy}(X,\vartheta,\upsilon)\,.
\end{equation}
The current $J_{\mathcal B}(X,\vartheta,\upsilon)$ is associated with the bosonic symmetries $SO(2,3)\times SU(4)$ and the current $J_{susy}(X,\vartheta,\upsilon)$ is associated with the 24 supersymmetries of the superbackground. The bosonic isometry current has the following generic structure
\begin{equation}\label{JB}
J_{\mathcal B}(X,\vartheta,\upsilon)=dX^MK_M(X)+J_1^A(X,\vartheta,\upsilon)\,K_A(X)+J^{[AB]}_2(X,\vartheta,\upsilon)\,K_A(X)K_B(X)\,,
\end{equation}
where $K_A(X)$ are the Killing vectors of $AdS_4\times CP^3$ and where $J_1$ and $J_2$ start at quadratic order in fermions.

The Lax connection is constructed using the pieces of the $OSp(6|4)$ Noether current and has the following structure
\begin{eqnarray}\label{LXT}
&\hspace{-20pt}\mathcal L(X,\vartheta,\upsilon)=\alpha_1 \,dX^MK_M(X)+ \alpha_2\,*J_{\mathcal B}+(\alpha_2)^2\,J_2+\alpha_1\alpha_2\,*J_2
+\alpha_2(-\beta_1\,J_{susy}+\beta_2\,*J_{susy})&\nonumber\\
&\hspace{-230pt}+\,\mathcal O(X,\vartheta,\upsilon^3)+\cdots\,, &
\end{eqnarray}
where $\cdots$ stands for terms which are higher order in $\upsilon$.

For generic motion of the string in the $\mathcal M_{10,32}$ superspace we have shown in \cite{Sorokin:2010wn} that this Lax connection has zero curvature at least up to the second order in fermions. When the fermions $\upsilon$ are put to zero this connection is related to the supercoset Lax connection \eqref{L} by an $OSp(6|4)$ gauge transformation. The construction of the higher order fermionic terms in the Lax connection \eqref{LXT} has turned out to be technically a rather involved problem. So far, the full Lax connection has been constructed only for the subsector of the complete theory which describes the string moving in $AdS_4$ and carrying eight fermionic excitations $\upsilon$ \cite{Sorokin:2010wn}, while the $CP^3$ embedding coordinates $y^{m'}$ and 24 fermionic modes $\vartheta$ are set to zero. However, since this is a subsector that is not reachable by the supercoset sigma--model, the proof of its integrability gives a major evidence for the classical integrability of the complete $AdS_4\times CP^3$ superstring. It would be useful, though, to find a systematic and geometrically grounded procedure for the construction of a Lax connection of the complete theory to all orders in the thirty two fermions.

\section{String instanton in $CP^3$}

Another interesting peculiarity of the $AdS_4\times CP^3$
superstring, which the $AdS_5\times S^5$ superstring does not have,
is the existence on $CP^3$ of string instantons. They are formed in the
Wick rotated theory by the string worldsheet wrapping a
topologically non--trivial two--cycle of $CP^3$. This two--cycle is
a $CP^1\simeq S^2$ corresponding to the closed K\"ahler two--form
$J_2$ on $CP^3$.

Let us first consider the purely bosonic $CP^3$--instanton solution of the Wick rotated string equations.  In this case the fermionic modes $\Theta$ are zero and the $AdS_4$ coordinates $x^m$ are worldsheet constants, while the $CP^3$ coordinates, given in terms of three complex variables $y^I$ $(I=1,2,3)$, are holomorphic (or anti--holomorphic) functions of the complex worldsheet coordinate $z=\tau+i\sigma$. The holomorphicity conditions
\begin{equation}\label{hol}
\frac{\partial y^I(z,\bar z)}{\partial \bar z}=0 \qquad \Rightarrow \qquad y^I=y^I(z)
\end{equation}
solve the string equations of motion and are just the $2d$ counterparts of the self--duality conditions satisfied e.g. by the $D=4$ Yang--Mills instantons. In this respect the string instanton in $CP^3$ is similar to the instantons in the two--dimensional chiral $CP^N$ sigma--models found back in the 70s \cite{Polyakov:1975yp,Golo:1978de}. The only difference is that in the case of the string one should check that also the Virasoro constraints are satisfied by eq. \eqref{hol}, which is indeed the case. One can further check that this instanton solution is 1/2 BPS, i.e. it preserve half of the supersymmetry of the background. This means that one can generate fermionic zero modes of the instanton by considering the variation of $\Theta$  under the supersymmetries which do not leave the purely bosonic solution invariant. In this way one can get twelve zero modes which play the role of goldstinos. But the instanton may also have fermionic zero modes that are not related to supersymmetries which it breaks. To check whether these modes do exist one should consider the fermionic part of the string action evaluated on the purely bosonic instanton solution (usually to the second order in fermions). In our case, in the conformal gauge and upon eliminating half of the 32 $\Theta$s using kappa--symmetry, we can write the instanton action in the following form
\begin{equation}\label{Iact}
S_I=n\,(\frac{R^2}{2\alpha'}-ia)+\frac{1}{\pi\alpha'}\int d^2\xi\,\det
{e}\,\left[i\,\vartheta\sigma^i\nabla_i\vartheta
-\frac{2}{R}\vartheta\vartheta
-2\big(i\,\upsilon\,\sigma^i\nabla_i\vartheta
-\frac{1}{R}\upsilon\upsilon\big)\right]\,,
\end{equation}
where the first term is the classical instanton action with $n$ being the instanton winding number and $R$ being the $CP^3$ radius. The second term is a constant axion contribution to the instanton action which may come from a non--zero expectation value of the NS--NS $B_2$--field proportional to the K\"ahler form $J_2$ on $CP^3$. Note that when such a $B_2$ flux is present, the $CFT_3$ dual of the string (and M) theory is the ABJ model \cite{Aharony:2008gk}.

The analysis of the Dirac equations for the fermions which follow from the action \eqref{Iact} shows \cite{Cagnazzo:2009zh} that the instanton does not have zero modes associated with the fermions $\upsilon$, while $\vartheta$ has twelve zero modes as we expected using the supersymmetry reasoning. It is interesting that these twelve zero modes are divided into eight and
four ones which have different geometrical and physical meaning. The eight massive fermionic zero modes are four copies of the two--component Killing spinor on the instanton surface $S^2$ and the four other fermionic modes are two copies of a massless chiral and anti-chiral fermion on $S^2$ electrically coupled to the electromagnetic potential created
on $S^2$ by a monopole placed in the center of $S^2$. The monopole potential arises as part of the $CP^3$ spin connection pulled--back to the instanton $S^2$.

The presence of the string instanton and its fermionic zero modes
may generate non--perturbative corrections to the string effective
action, which may affect its properties and if so should be taken
into account in studying, \emph{e.g.} the $AdS_4/CFT_3$
correspondence.  For instance, the instantons may, probably, contribute to the
worldsheet S--matrix and/or to energies of a semiclassical string.
Note that quite recently Drukker, Mari\~no and Putrov \cite{Drukker:2010nc} found contributions coming from worldsheet instantons to the partition function and Wilson loop observables computed in a matrix model description of ABJ(M) theory.

\section{Conclusion}\label{summary}

In this contribution we have discussed peculiar features of the string theory in the $AdS_4\times CP^3$ superbackground regarding the structure of its action, classical integrability and the worldsheet instantons on $CP^3$.
Other aspects of the $AdS_4\times CP^3$ superstring theory, in particular, different kappa--symmetry gauge fixings simplifying its action, Penrose and other limits have been studied \emph{e.g.} in \cite{Uvarov:2008yi,Uvarov:2009hf,Uvarov:2009nk,Astolfi:2009qh,Grignani:2009ny,Bykov:2010tv}.

In conclusion we would like to mention one more issue of this theory, the so called fermionic T--duality of the Green--Schwarz action. In the case of the $AdS_5\times S^5$ superstring there is a worldsheet T--duality of bosonic string coordinates along the Minkowski boundary of $AdS_5$ \cite{Kallosh:1998ji,Ricci:2007eq} accompanied by a dual transformation of eight (complex) fermionic coordinates (associated with translational isometries)
\cite{Berkovits:2008ic,Beisert:2008iq} which brings the $AdS_5\times S^5$ superstring action to itself but in a different kappa--symmetry gauge. It has been shown that this self--duality of the $AdS_5\times S^5$ superstring is related to earlier observed dual conformal symmetry of maximally helicity violating amplitudes of the ${\mathcal N}=4$ super--Yang--Mills theory and to the relation between gluon scattering amplitudes and Wilson loops at strong and weak coupling \cite{Alday:2007hr,Drummond:2007aua,Drummond:2007cf,Brandhuber:2007yx,Drummond:2008vq,Komargodski:2008wa,Ricci:2007eq,Berkovits:2008ic,Beisert:2008iq}.
 In the context of the $AdS_4/CFT_3$ correspondence, manifestations of the dual superconformal symmetry in the ABJM model have been found in \cite{Bargheer:2010hn,Henn:2010ps,Lee:2010du,Huang:2010qy,Gang:2010gy}. The bosonic T--duality of the $AdS_4\times CP^3$ superstring action has been performed in \cite{Adam:2009kt,Grassi:2009yj}, however the dualization of string fermionic coordinates encountered problems \cite{Adam:2009kt,Grassi:2009yj,Hao:2009hw,Adam:2010hh,Bakhmatov:2010fp,Dekel:2011qw} caused, in particular, by a singularity of the matrix transforming original variables to the dual ones. So the worldsheet self--duality of the $AdS_4\times CP^3$ superstring still remains an important open issue whose solution would shed light on the holographic nature of the dual superconformal symmetry in the ABJM model and would give a better understanding of the interrelation between the integrable structures on both sides of the $AdS_4/CFT_3$ duality. Finally, let us note that, as has been pointed out in \cite{Bergman:2009zh} (see also \cite{McLoughlin:2008he}), the tension of the string in $AdS_4\times CP^3$ should acquire a worldsheet two--loop correction which results in a shift of the 't Hooft coupling $\lambda$. It would be of interest to compute this correction directly using the $AdS_4\times CP^3$ superstring action.

\section*{Acknowledgements}
 This work was partially supported by the INFN Special
Initiative TV12. D.S. was also partially supported by an Excellence
Grant of Fondazione Cariparo (Padova) and the grant FIS2008-1980 of
the Spanish Ministry of Science and Innovation. The  research of L.W. is supported in part by
NSF grants PHY-0555575 and PHY-0906222.

\if{0}
\bibliography{strings,BLG}
\bibliographystyle{utphys}
\end{document}
\fi

\providecommand{\href}[2]{#2}\begingroup\raggedright\endgroup

\end{document}